\documentclass[%
 reprint,
 superscriptaddress,
 amsmath,amssymb,
 aps,
 prl,
]{revtex4-2}
\usepackage{xcolor}
\usepackage{lipsum} % for dummy text only

\usepackage{appendix}
\usepackage{braket}
\usepackage{float} 
\usepackage{chemformula}
\usepackage{graphicx}% Include figure files
\usepackage{dcolumn}% Align table columns on decimal point
\usepackage{bm}% bold math
\usepackage{setspace}
\usepackage{physics}
\usepackage{braket}
\usepackage{graphicx}% Include figure files
\usepackage{dcolumn}% Align table columns on decimal point
\usepackage{bm}% bold math
\usepackage[colorlinks=true, citecolor=blue, linkcolor=blue, urlcolor=blue]{hyperref}
\usepackage{indentfirst}

\usepackage[thinc]{esdiff}

\def \ETH{Institute for Theoretical Physics, ETH Z\"urich, CH-8093 Z\"urich, Switzerland}
\def \Harvard{Lyman Laboratory, Department of Physics, Harvard University, Cambridge, MA 02138, USA}
\def \EPFL{Institute of Physics, Ecole Polytechnique Fédérale de Lausanne, Lausanne, Switzerland.}
\def \Geneva{Department of Quantum Matter Physics, University of Geneva, Quai Ernest-Ansermet 24, 1211 Geneva, Switzerland}
\def \Kaiserslautern{Physics Department and Research Center OPTIMAS, University
of Kaiserslautern-Landau, D-67663, Kaiserslautern, Germany}
\begin{document}

\title{Dynamical Instabilities of Strongly Interacting \\ Ultracold Fermions in an Optical Cavity}

\author{Filip Marijanovi\'c}
\thanks{These authors contributed equally to this work}
\affiliation{\ETH}

\author{Sambuddha Chattopadhyay}
\thanks{These authors contributed equally to this work}
\affiliation{\ETH}
\affiliation{\Harvard}

\author{Luka Skolc}
\affiliation{\ETH}

\author{Timo~Zwettler}
\affiliation{\EPFL}

\author{Catalin-Mihai~Halati}
\affiliation{\Geneva}

\author{Simon B.~J\"ager }
\affiliation{\Kaiserslautern}

\author{Thierry~Giamarchi}
\affiliation{\Geneva}

\author{Jean-Philippe~Brantut}
\affiliation{\EPFL}

\author{Eugene Demler}
\affiliation{\ETH}

\date{\today}

\begin{abstract}
Recent quench experiments on ultra cold fermions in optical cavities provide a clean platform for studying how long-range interactions between fermions structure their dynamics.  Motivated by these experiments, we provide a theoretical analysis of the dynamical instabilities that lead to the formation of superradiance as the hybrid system is driven across the self-organization transition. We compute the rate at which order forms and quantify the fluctuations of the pre-quench state which seed the instability. Our results quantitatively match existing experiments on free fermions and make predictions for quench experiments involving near unitary fermi gases coupled to an optical cavity. Our work suggests that the non-local nature of the photon-mediated interactions between fermions generates ordering dynamics that are qualitatively different than those observed in short-range interacting systems.
\end{abstract}
\maketitle

\textit{\textbf{Introduction} --- }
How fast does order grow in a fermionic many-body system as it is quenched across a quantum phase transition? The answer to this question---demanding replies from contexts as varied as the cosmological \cite{kibble} to the microscopic \cite{Zurek,stamper-kurn,ketterle1, ketterle2, pekker}---is constrained by \textit{locality}. As a concrete example, consider experiments performed on ultracold fermions in which Feshbach mediated interactions \cite{ketterle1, ketterle2} were rapidly quenched. Here the ``speed-limit" to the post-quench growth rate of both superconducting  and ferromagnetic order in an interacting fermi gas is set by the $E_{\rm F}$, the Fermi energy \cite{pekker}. What complications arise, then, when we ask the question in a system with non-local interactions?

Recent experiments  \cite{shanghai_superradiance, dwo, shanghai_dynamics, lausanne_dynamics} investigating the quench dynamics of ultracold fermions in an optical cavity present an ideal laboratory for such questions \cite{cmh1,cmh2, cmh3, cmh4, cmh5,cmh6,cmh7}. Leveraging techniques used to realize global range interactions between \textit{bosons}  \cite{bosonTheo1, bosonTheo2, bosonTheo3, vuletic, tilman} in these experiments, strong cavity-mediated all-to-all interactions are engineered between \textit{fermions} by loading quantum degenerate atoms into a cavity and driving them with a transverse pump. Under sufficiently strong driving, the initially homogeneous gas self-organizes into a density wave order, which concomitantly supports the superradiant build-up of cavity photons \cite{tilman, cavity_rmp_old, zhai_superradiance, keeling_superradiance,
cavity_review_new}. In experiments, the transverse pump power is rapidly increased, quenching the system from the normal to ordered phase. Given the controllably dissipative nature of the cavity, photons leak out of the cavity in the superradiant phase \cite{shanghai_dynamics, lausanne_dynamics}. Remarkably, the leaked photons provide a high-resolution \textit{in vivo} probe, allowing experiments to (weakly) monitor the dynamics of superradiant self-organization.

   \begin{figure}[H]
        \centering
        \includegraphics[width = 0.96 \linewidth]{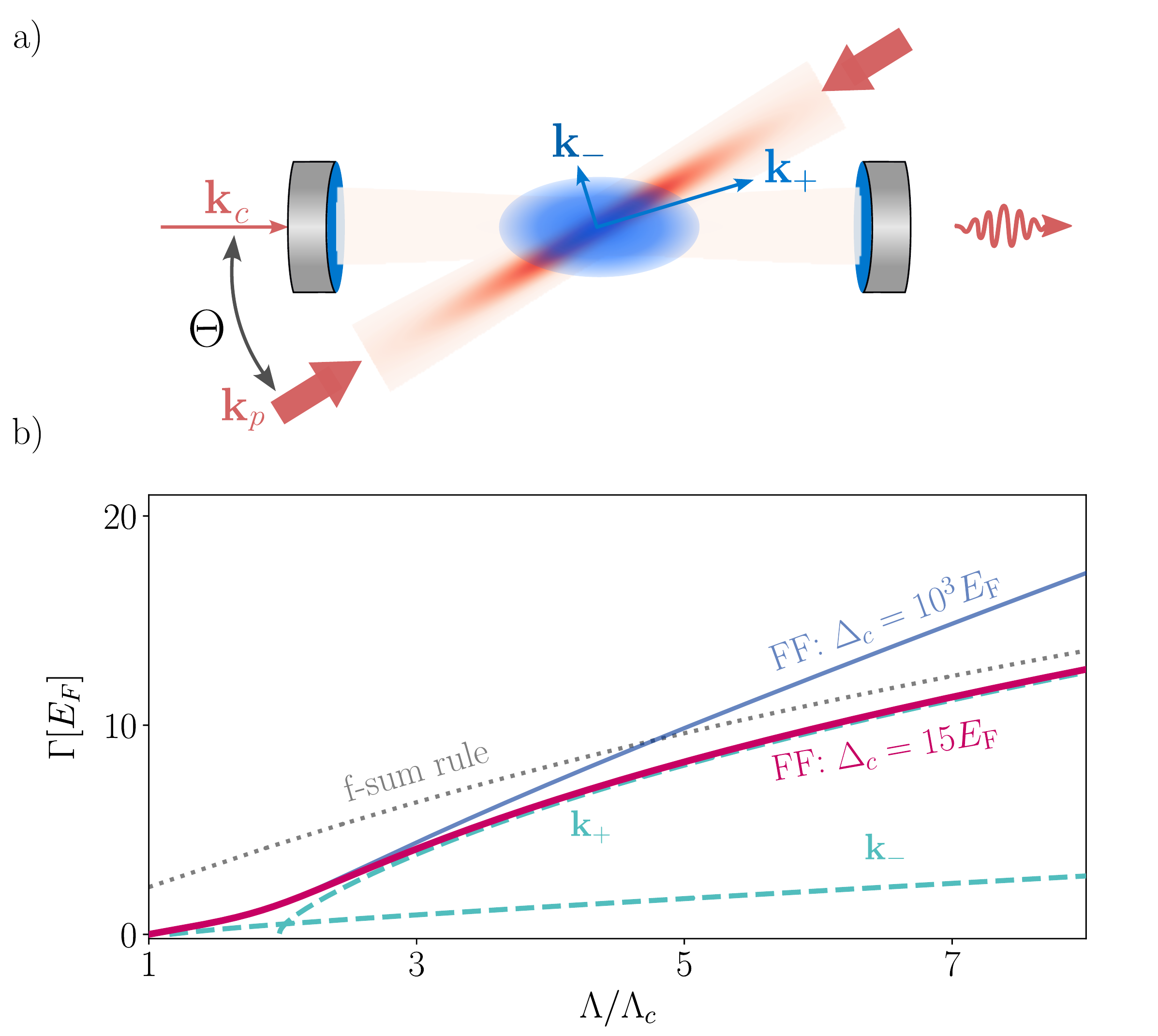}
        \caption{\textbf{Instability Rate} (a) Experimental schematic-- pump incoming at an angle $\Theta$ with respect to the cavity. For powers above the phase transition the system is superradiant and the fermionic density orders at two wavevectors $\textbf{k}_\pm$, given by cavity and pump wavevectors $\textbf{k}_{c/p}$ respectively. (b) The instability growth rate $\Gamma$ for 3D free fermions (FF) in units of the Fermi Energy $E_{\rm F}$  as a function of the reduced quenched photon-fermion coupling strength $\Lambda/\Lambda_c$, plotted for two representative pump-cavity detunings $\Delta_c = 15 E_{\rm F}$ and $\Delta_c = 10^3 E_{\rm F}$. Dashed lines show contributions from the two ordering wavevectors $|\textbf{k}_+| = 2.24k_{\rm F}$ and $|\textbf{k}_-| = 0.36 k_{\rm F }$ for $\Delta_c=15E_{\rm F}$. The strong quench behavior is constrained by the f-sum rule (dotted).}
        \label{fig:Fig1}
    \end{figure}
    
In this \textit{Letter}, we calculate how fast the intertwined density and superradiant orders develop in the system after a quench, quantifying both the initial fluctuations which seed the instability as well as the rate of the order's exponential growth. We make direct contact with experiments, matching growth rate measurements for dissipative free-fermions coupled to an optical cavity and predicting rates for cavity quench experiments performed on strongly interacting fermions near unitarity. We conclude by extrapolating our theoretical results to regimes beyond current experiments, commenting on how non-local interactions alter ordering in fermionic many-body systems.
    
\textit{\textbf{Model} ---} We consider fermions in the lower energy internal state of a two-component, quantum degenerate fermi gas, dispersively coupled to an optical cavity with wavevector $\textbf{k}_c$ which is transversely driven by a retro-reflected laser with wavevector $\textbf{k}_p$ \cite{dwo, shanghai_superradiance, shanghai_dynamics, lausanne_dynamics} (see Fig. \ref{fig:Fig1}). In the rotating frame of the pump, the effective Hamiltonian is given by \cite{dwo}
    \begin{equation}
        H =H_{\rm F} + \frac{\Delta_c}{2} (x^2 + p^2) + H_{\rm int}
        \label{eq:hamiltonian}
    \end{equation}
    where $H_{\rm F}$ is a generic fermionic Hamiltonian (containing, e.g. local, Feshbach-mediated pairing interactions), $\Delta_c$ is the \textit{effective} photon detuning (accounting for the collective dispersive shift from the atoms) and $x,p$ are photon quadratures. $H_{\rm int}$ is the fermion-cavity coupling given by
    \begin{equation}
        H_{\rm int} =  \sqrt{N} \eta(t) x \sum_{Q} (\rho_{Q}+\rho_{-Q})
        \label{eq:interacting_hamiltonian}
    \end{equation}
where the fermionic density operator at wavevector $Q \in \{\textbf{k}_\pm = \textbf{k}_c \pm \textbf{k}_p\}$ is written as $\rho_{Q} = \frac{1}{\sqrt{N}} \sum_{p, \sigma} c^\dag_{p+Q,\sigma}c_{p,\sigma}$ and $c^\dag_{p\sigma}(c_{p\sigma})$ are fermionic creation (annihilation) operators that create (annihilate) a fermion of momentum $p$ and spin $\sigma$. The interaction between fermionic density modulations at wavevectors $\textbf{k}_\pm$ and cavity photons arises from the AC stark shift due to the interference potential generated at $\textbf{k}_\pm$ between the retro-reflected beam and the cavity field. The coupling strength $\eta(t)$ can be tuned by modulating the  power of the transverse driving field, as is done in the quench experiments. We ignore the pump induced AC-Stark shift which does not play a role in self-organization. We also, temporarily, neglect cavity dissipation. In the absence of dissipation, the steady-state of the system is superradiant above $\eta_c = \ \frac{1}{2} (\Delta_c / (\sum_Q \chi(Q) N))^{\frac{1}{2}}$ \cite{zhai_superradiance, dwo} -- where $\chi(Q) \equiv \chi(Q, \omega=0)$ is the interaction dependent static atomic density susceptibility at wavevector $Q$ -- and normal (i.e., no cavity photons, homogenous fermi gas) below it. As a reminder, within linear response theory, for a system externally driven by $H^{\rm ext} = \phi e^{i\omega t} \rho_{-Q}$, we introduce the atomic susceptibility is given as $\chi(Q,\omega) = \expval{\rho_Q}/\phi$ ~\cite{dupuis2023field}.

\textit{\textbf{Instability Rate} ---} Having summarized the steady-state phases of our system, we now consider a quench of the light-matter interaction $\eta(t)$ from $0$ to $\Lambda \ge \eta_C$ at $t=0$. After a brief transient regime, density order rapidly forms, growing as $\rho_Q(t) \simeq \rho_{Q,0} e^{\Gamma t}$ while photons simultaneously proliferate exponentially in the cavity, evolving as $n(t) \simeq n_0 e^{2 \Gamma t}$ in the unstable regime. We calculate the growth rate $\Gamma$ by studying the dynamics of the unstable, polaritonic collective mode within a linear approximation. Specifically, we begin by writing down the Heisenberg equations of motion for the photonic quadratures in frequency space: $i \omega x = \Delta_c p;\,  i \omega p = -\Delta_c x - \sqrt{N} \eta(t) \sum_{Q, \sigma} \rho_{Q,\sigma}$. Next, we recognize that $H_{\textrm{int}}$ has precisely the form of the \textit{gedanken} probe field used to derive the fermionic density susceptibility within linear response. Leveraging this association, we write $\rho_{Q,\sigma} = \sqrt{N} \eta(t) \chi(Q,\omega) x$, where $\chi(Q,\omega)$ is the (dynamical) atomic density susceptibility at frequency $\omega$ and wavevector $Q$. We model the quench by taking $\eta(t=0^{+}) = \Lambda$ and $\chi(Q, \omega)$ of the initially uncoupled, possibly interacting, ultracold fermi gas. Taken together, we find a closed set of equations that allow us to find an expression for the frequency of the polaritonic collective mode arising from the hybridization between the cavity photons and fermionic density modes at $Q$ 
    \begin{equation}
        \omega^2 = \Delta_c^2 - 4 \Lambda^2 N \Delta_c  \sum_{Q} \chi(Q,\omega).
    \label{eq:rate_equation}
    \end{equation}
Our self-consistency equation admits purely unstable, imaginary solutions $\omega = \pm i \Gamma$ beyond a critical coupling $\Lambda_c$ -- see the Supplement \cite{SM} for details of the solution. Using the spectral representation, we can recast the self-consistent equation for $\Gamma$ as
\begin{equation}
   1 + \Gamma^2 \Delta_c^{-2} = \frac{4\Lambda^2 N}{\Delta_c} \sum_Q   \int_{-\infty}^{\infty} \frac{d\omega}{\pi}\frac{\omega \Im \chi(Q,\omega)}{\omega^2+\Gamma^2 }.
    \label{eq:rate}
\end{equation}

\begin{figure}[t]
\centering
\includegraphics[width = 0.96 \linewidth]{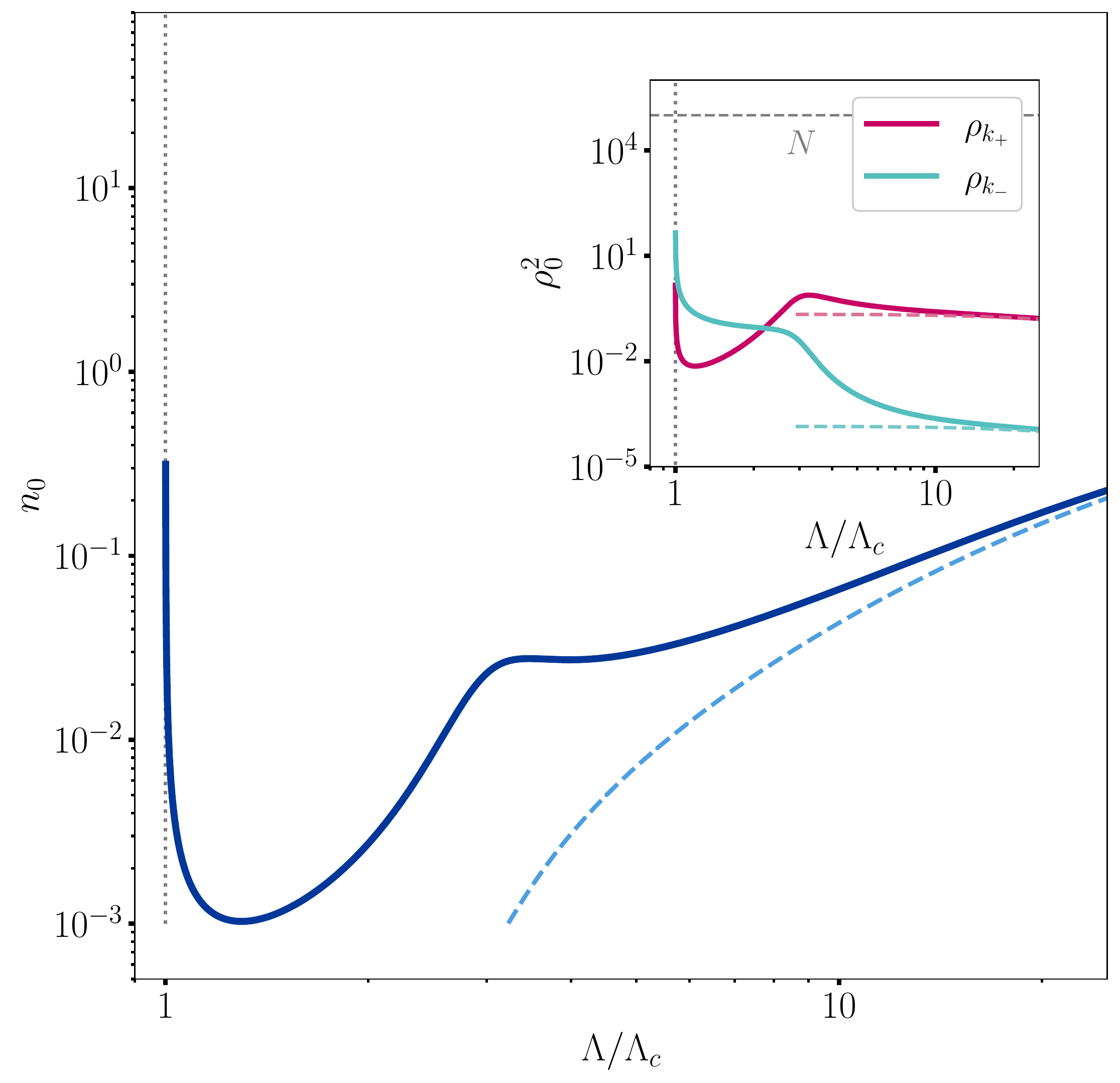}
\caption{\textbf{Instability Seeded By Density Fluctuations} The photon number pre-factor $n_0$ as a function of the light-matter coupling $\Lambda$ for $\Delta_c = 175E_{\rm F}$ and ordering wave vectors $|\textbf{k}_{+}| = 3.7k_{\rm F}$, $|\textbf{k}_-| = 0.59 k_{\rm F}$ for a 3D unitary fermi gas. The dashed lines show the asymptotic behaviours for deep quenches. Inset: The dependence of the density ordering pre-factor on $\Lambda$ for the $\textbf{k}_{\pm}$ modes, in the same parameter regime. The gray horizontal line shows an estimate to the saturation value of density fluctuations.}

\label{fig:Fig2}
\end{figure}

We underscore that---within a linear-response approximation---this equation is completely general, allowing for \textit{quantitative} comparison with experiments involving myriad atomic media (e.g. fermions, bosons; free or interacting; homogenous or with a harmonic trap; at zero or finite temperature; in a single-mode or multi-mode cavity). 

Asymptotic insight into \eqref{eq:rate} can be gained by considering the sum rules imposed on $\Im \chi(Q, \omega)$. The low frequency behavior of $\Im \chi(Q, \omega) $ is constrained by a generalized ``finite-$Q$" compressibility rule  arising from the Kramers-Kronig relations on $\Im \chi(Q, \omega )$: $\int \mathrm{d}\omega \omega^{-1} \Im \chi(Q,\omega)=\pi \chi(Q)$. This compressibility sum rule can be used to find the critical quench strength $\Lambda_c =\frac{1}{2} (\Delta_c / (\sum_Q \chi(Q) N))^{\frac{1}{2}} = \eta_c$, as expected from a naïve Ginzburg-Landau free energy picture (see Fig.~\ref{fig:Fig1}). The high frequency behavior of $\Im \chi(Q, \omega)$, which dictates the behavior of $\Gamma$ for deeper quenches, is constrained by the $f$-sum rule: $\int  \mathrm{d}\omega \omega \Im \chi(Q,\omega)  = \pi Q^2/m$. For deeper quenches $\Lambda \gg \Lambda_c$, for which $\Gamma$ exceeds the effective bandwidth of the fermions, our self consistency equation reduces to $\Gamma^2\big(1 + \Gamma^2 \Delta_c^{-2}\big) = \frac{8\Lambda^2 N E_\textrm{R}}{\Delta_c}$, where $E_\textrm{R} = \frac{\textbf{k}_-^2+\textbf{k}_+^2}{2 m}$. For stronger quenches, this implies a universal behavior of $\Gamma$, invariant to microscopic details. If the cavity field and transverse drive are sufficiently non-orthogonal, $\textbf{k}_+ \gg \textbf{k}_-$  \cite{lausanne_dynamics}. As a result, $\chi(\textbf{k}_-) \gg \chi(\textbf{k}_+)$, suggesting that for critical quenches, the collective instability is driven by the formation of density order along $\textbf{k}_-$. Per contrast, for deeper quenches the instability arising from $\textbf{k}_+$ dominates the one arising from $\textbf{k}_-$. Taken together, for experiments in which $\textbf{k}_+ \gg \textbf{k}_-$, there is a dual nature to the instability, exhibiting an inflection point in the rate at larger coupling strengths, as shown in Fig.~\ref{fig:Fig1}.

Naively, \eqref{eq:rate} can be expanded around the critical point to obtain the critical exponent associated with $\Gamma$. However, if the support of $\Im \chi(Q,\omega)$ extends to $\omega = 0$, $\Gamma$ is not analytic around the critical point. Accounting for this carefully, we find that around the critical point the integral in \eqref{eq:rate} can be expressed to leading order as $\Gamma \arctan(E_{\textrm F}/\Gamma) \approx \frac{\pi}{2} |\Gamma|$, yielding that $\Gamma \propto \big(\Lambda - \Lambda_c\big)$, in contrast with the dissipationless Dicke model where $\Gamma \propto \big(\Lambda - \Lambda_c\big)^\frac{1}{2}$ \cite{dynamicsDicke}. The traditional Dicke response is recovered for a system which has a gap in $\Im \chi(Q,\omega)$, usually associated with a recoil energy in the modelling of many-body cavity-QED experiments. While such a gap may arise for strictly non-interacting fermions in low dimensions or if all of the ordering wavevectors $Q > 2k_{\rm F}$, such provisos are not applicable to current experiments \cite{shanghai_dynamics, lausanne_dynamics}. 

\textit{\textbf{Seeding the Instability} ---} Having computed the growth rate $\Gamma$, it is natural to ask: ``How long does the unstable regime last?" To answer this, we calculate $n_0$ and $\rho_{Q,0}$---pre-factors for the exponentially growing photon number and the density order in the superradiant instability---and compare them to estimates for $n_{\rm sat}$ and $\rho_{\rm Q, sat}$ where we expect non-linear saturation effects to arise. To calculate pre-factors $n_0$ and $\rho_{Q,0}$, we note that---excluding exogenous sources of photons---the instability is dominantly seeded by the density fluctuations of pre-ordered fermi gas \cite{SM}.  

We quantify how these density fluctuations seed the instability by computing their power spectral density at the instability frequency $\Gamma$. To do so, we turn to the Laplace formalism which allows us to propagate the linearized Heisenberg evolution arising from (\ref{eq:hamiltonian}), even in the presence of an exponentially growing mode. In the Supplement \cite{SM} we provide both a derivation of the pre-factor by leveraging the Kubo formula and an explicit calculation of the pre-factor for free-fermions---we briefly sketch our general derivation here. Within the Laplace formalism, the photonic equations of motion become: $ s x(s) - x(t = 0) = -i\Delta_c p(s) $ and $s p(s) - p(t=0) = i\Delta_c x(s) + i \sqrt{N} \Lambda \sum_Q \rho_Q(s)$. For the fermionic evolution we use the Kubo formula which, upon Laplace transformation, reads: $\rho_Q(s) = \sqrt{N} \Lambda \chi(Q,s) x(s) + \rho_Q^0(s)$, where $\rho_Q^0(s)$ is the Laplace transform of the interaction picture evolution of $\rho_Q(t=0) = e^{i H_F t} \rho_Q(t=0) e^{-i H_F t}$. Combining the equations and expressing $\rho^0_Q$ via the Lehmann representation, we find that $n_0$, the pre-factor for photons, is given by

\begin{widetext}
\begin{equation}
    n_0 =  \left. \frac{1}{8}\left(\frac{1}{\Gamma^2} + \frac{1}{\Delta_c^2}\right)\frac{\Lambda^2 \Delta_c^3}{ \Lambda_c^2 \sum_Q \chi(Q)}  \sum_Q \int_0^\infty d\omega \frac{\Im \chi(Q, \omega)}{\omega^2 + \Gamma^2} \middle/ \left(1 + \frac{\Lambda^2 \Delta_c^2}{ \Lambda_c^2 \sum_Q \chi(Q)}\sum_Q \int_{-\infty}^\infty  d\omega \frac{\Im \chi(Q, \omega) \omega}{(\omega^2 + \Gamma^2)^2}\right)^2 \right.
\label{eq:seed}
\end{equation}
\end{widetext}

and the density ordering pre-factor $\rho_{Q,0} = \frac{1}{\sqrt{2}}\frac{\Lambda}{\Lambda_c}\sqrt{\frac{\Delta_c^3}{\sum_Q \chi(Q)\left(\Gamma^2+\Delta_c^2\right)} }\chi(Q,i \Gamma) \sqrt{n_0}$. For near critical quenches, the pre-factors diverge weakly with $n_0$ and $\rho_{Q,0} \sim \log(E_{\rm F}/\Gamma)$, whereas for deep quenches, $ n_0 = \frac{1}{8} (2 + \frac{\Gamma^2}{\Delta_c^2} + \frac{\Delta_c^2}{\Gamma^2}) (\frac{\Gamma^2}{2 \Gamma^2+ {\Delta_c^2}})^2 \frac{\Delta_c}{2 E_{\rm R}} \sum_Q S(Q)  $, $\rho_{Q,0} = \frac{1}{2\sqrt{2}} \frac{E_Q}{E_R} \frac{\Gamma^2 + \Delta_c^2}{2\Delta_c^2 + \Gamma^2} $, where $S(Q)$ is the static structure factor of the atomic gas at wavevector $Q$ and $E_Q$ is the recoil energy at momentum $Q$. The log divergence for near critical quenches arises from the fact that the support of $\Im \chi(Q,\omega)$ extends to $\omega = 0$.

Our asymptotics suggest the non-monotonic behavior of $n_0$ and $\rho_{Q,0}$ as a function of $\Lambda$ as shown in Fig.~\ref{fig:Fig2}, which exhibits a minimum within the experimental regime---we now briefly describe the physical origins of the divergences in each limit. For near-critical quenches, we expect that dynamical fluctuations of the order parameter in the Ginzburg-Landau potential with a vanishing quadratic term diverge as the frequency goes to 0, implying a divergent pre-factor as a function of $\Gamma$. In contrast, for deep quenches, the growth rate is so fast that the instability is only sensitive to the static fluctuations of the gas described by $S(Q)$. At the same time, the character of the unstable polaritonic mode becomes more density-like. Thus, the pre-factor---which carriers a projection onto the unstable mode---picks up a factor that scales with $\Lambda$. 

As the polaritonic mode grows and photons proliferate in the cavity, density modulations deplete fermions in the nodes of the waves along $Q$. As $\rho_Q$ approaches $\mathcal{O}(\sqrt{N})$, non-linearities arise to penalize further depletion of the fermionic density, leading to saturation. While a more careful analysis of this saturation is required, a strict upper bound on $\rho_{Q,\textrm{sat}}^2$ is set by $N$ ($\sim10^5$ in experiments), a full condensation of the density mode along $Q$. 

Juxtaposing this upper bound on $\rho_{Q,\textrm{sat}}^2$ with $\rho_{Q,0}^2$, as plotted in the inset of Fig.~\ref{fig:Fig2}, reveals a dynamic range of at least 5 orders of magnitude immediately away from criticality. Thus, non-linear saturation effects are weak and the (linear) instability should persist for multiple decades of exponential growth. The inset of Fig.~\ref{fig:Fig2} also reflects that $\rho_{\textbf{k}_+,0}/ \rho_{\textbf{k}_-, 0} = \chi(\textbf{k}_+,i \Gamma)/ \chi(\textbf{k}_-,i \Gamma)$ which implies that while for shallower quenches the pre-factor for the $\textbf{k}_-$ mode is larger than that of the $\textbf{k}_+$ mode, the situation is reversed for deeper quenches. Thus, within our saturation mechanism, for near critical quenches, the $\textbf{k}_-$ mode should saturate first while for deeper quenches, the $\textbf{k}_+$ instability depletes the fermions faster.

%\footnote{We remark in passing that one may worry that as for both $\Lambda \to \Lambda_c$ and $\Lambda/\Lambda_c \to \infty$ $\rho_{Q,0}\gg \rho_{\rm sat}$, such constraints may strongly circumscribe the parameter regime where a meaningful instability can arise. We note first that the former anxiety is relevant only in a small vicinity around $\Lambda_c$, given the weak, logarithmic nature of the divergence as $\Gamma \to 0$. Furthermore, reaching the large $\Gamma$ regime where the pre-factor becomes comparable to $\rho_{\rm sat}$ requires reaching experimentally unreasonable powers. At such powers, one expects that exogenous experimental details, ranging from increased three-body losses to density non-linearities arising from the modulation due to the AC stark shift from the transverse drive, to become relevant. }. 

\textit{\textbf{Experimental implications} ---}While the phenomenology discussed so far is experimentally feasible, experiments typically operate in the regime where $\Delta_c/E_{\rm F} \sim 100-300$ and for quenches up to $\Lambda \sim 5 \Lambda_c$ \cite{lausanne_dynamics, shanghai_dynamics}, probing primarily the region around criticality. The first condition implies that dynamics of the cavity photon are irrelevant and the instability can be captured by studying quenches of fermions interacting all-to-all with cavity photon mediated interactions. This in turn means that as a function of $\Lambda / \Lambda_c$, $\Gamma$ does not depend on $\Delta_c$, a fact that is captured by the negligibility of the $\Gamma^2 \Delta_c^{-2}$ term in (\ref{eq:rate}) in the experimental regime. The second experimental circumscription implies that the scale of the instability rate is set by $E_{\rm R}$, highlighting the fermionic character of the instability. 

\begin{figure}[t]
        \centering
        \includegraphics[width = 0.96 \linewidth]{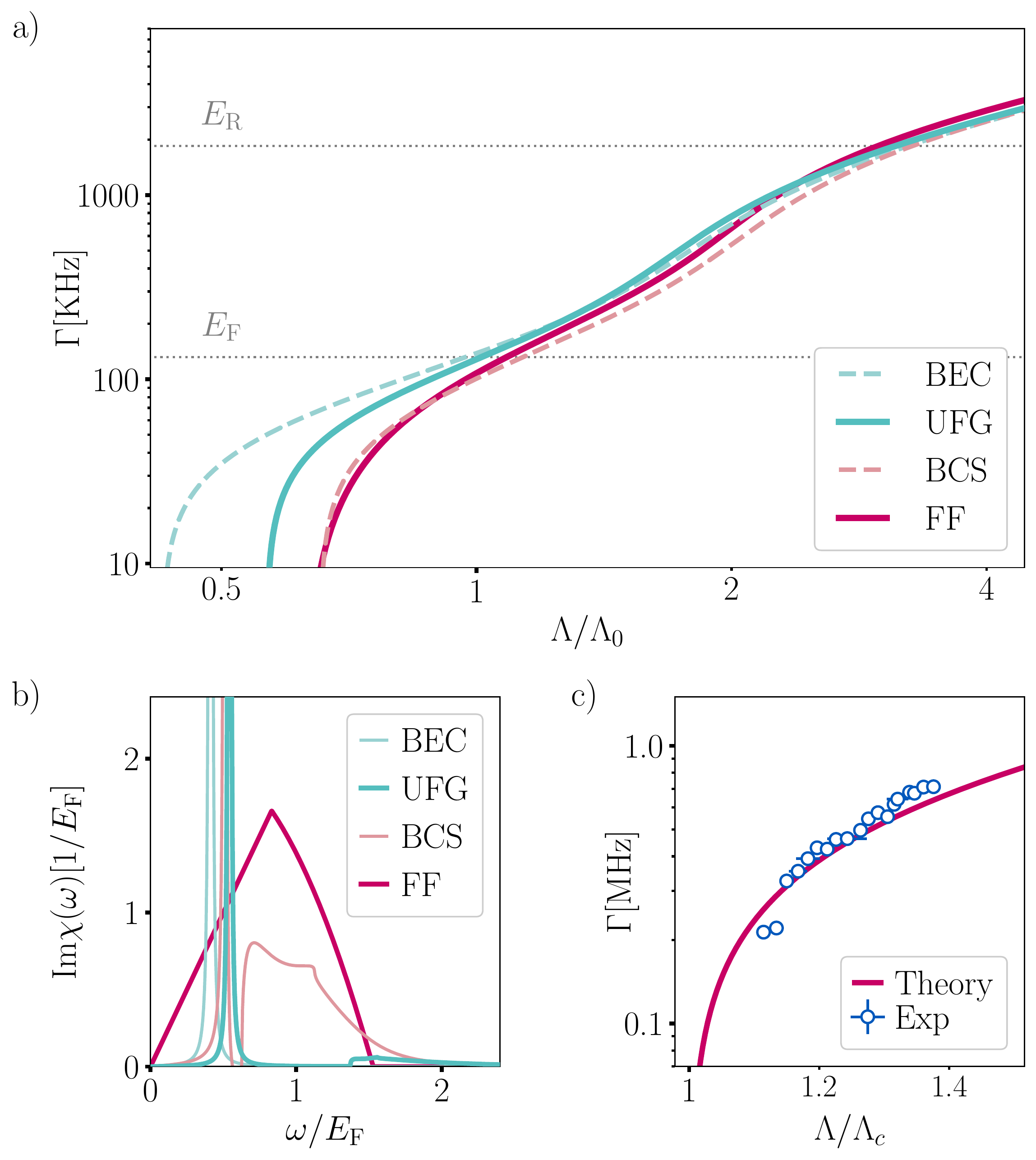}
        \caption{\textbf{Interactions and experiments} (a) Instability growth rate $\Gamma$ as a function of effective coupling in 3D for different scattering lengths across the BEC-BCS crossover $1/k_{\rm F}a = \{-0.7,0,1.1\}$ and for free fermions. The quench strength $\Lambda$ is measured in units of $\Lambda_0 = \sqrt{\frac{\Delta_c E_{\rm F}}{N}}$. 
        $\Delta_c = 100 E_{\rm F}$, with $\textbf{k}_+ = 3.7 k_{\rm F}$ and $\textbf{k}_- = 0.59 k_{\rm F}$, modelling the experimental setup \cite{lausanne_dynamics}. (b) Low frequency behavior of $\Im{\chi(\textbf{k}_-, \omega)}$ across the BEC-BCS crossover, within the RPA approximation. Interactions induce spectral features that are absent for free fermions, including a Bogolyubov-Anderson mode and quasi-particle continua. These features result in quantitative differences in the rates, as seen in the previous panel. (c) Comparison of the rate, calculated for free fermions including cavity dissipation and trap averaging, with experimental data (See Supplement \ref{sec:Exp-comp} for details) \cite{shanghai_dynamics}.}
        \label{fig:Fig3} 
        
\end{figure}

Experiments also feature finite dissipation rate arising from the leakage of photons from the cavity at a rate $\kappa$. We can capture this straightforwardly by taking $\omega \to \omega-i \kappa$ in \eqref{eq:rate_equation} on the left hand side but \textit{not} on the right \cite{SM}. In Fig.~\ref{fig:Fig3}a, we show results comparing quenches of different interacting systems across BEC-BCS crossover from $\frac{1}{k_F a} = -0.7-1.1$, corresponding to experiments performed on near unitary fermi gases \cite{dwo, lausanne_dynamics}. As shown in Fig.~\ref{fig:Fig3}b, distinctions between these rates arise from qualitative differences in the dynamical susceptibility of the atomic gases---as compared to free fermions, interactions lead to a sharp Bogolyubov-Anderson mode at $\omega \sim c_s \textbf{k}_-$ where $c_s$ is the sound velocity in the atomic gas. The differences in the $\Lambda_c$ across the interacting regime arise then from the differences of the sound velocity across the BEC-BCS transition. In Fig.~\ref{fig:Fig3}c, we make an explicit comparison to experiments performed on dissipative, harmonically trapped, free-fermions \cite{shanghai_dynamics}. We find good agreement with experimental measurements upon accounting for harmonic trap induced density inhomogeneities \cite{SM}.

\textit{\textbf{Discussion} ---} In our work we articulate a linear theory that both captures the rate at which order grows after a quench into the self-organized phase and quantifies the fluctuations which seed the instability. Our theory is quantitatively accurate in capturing existing experimental measurements of the instability rate and makes  predictions for future experiments studying quenches in optical cavities where the atomic gas is strongly interacting. Moreover, the analytical tractability of the present setting in which a couple discrete unstable collective modes arise allows us to elucidate various non-trivial aspects of fermionic instabilities---e.g., the changing character of the coupled unstable modes that seed and saturate the order parameter growth and the rigorous use of a Laplace analysis to quantify pre-existing fluctuations. 
We expect that the careful, experimentally verifiable theoretical considerations of such matters in this simpler context will facilitate the analysis of more complex settings in which there are a plethora of unstable modes. 

Extensions of our analysis may be useful in interpreting experiments which seek to optically control complex materials. While experiments in this burgeoning field have probed a wide variety of ordering dynamics---ranging from the growth of non-equilibrium charge density order \cite{CDW,Kogar_20} to the photo-induction of metastable ferroelectricity \cite{li_terahertz_2019, Nova_19} and superconductivity far above $T_c$ \cite{sc_1, sc_2, sc_3}---theoretical modelling of such dynamical phenomena remain limited in either scope or microscopic detail due to their computational intractability \cite{theo1, theo2, theo3, theo4}. Extensions of the formalism developed here---along with a more sophisticated understanding of non-linear and dissipative saturating effects---could pave the way for a tractable theoretical framework for understanding how to stabilize non-thermal states of matter in solid-state settings.

Given the success of our theory in capturing experimental measurements, we conclude by highlighting a startling discrepancy between the growth rates that we uncover herein and the growth rates of short-range interacting systems where $E_{\rm F}$ sets a local speed limit on the rate at which the superconducting or ferromagnetic order parameters can grow. For deep quenches in our setting---restricting our discussion to the vast regime of quenches where $\Gamma \ll \Delta_c$---(\ref{eq:rate}) does not provide a bound on the rate, allowing it to \textit{far exceed} all fermionic scales, including $E_{\rm F}$ and $E_{\rm R}$. 

To gain physical insight into this surprise that our formalism uncovers, we can first consider why local limits arise in short-range interacting systems where $E_{\rm F}$ cuts off the scale of the instability rate. If the interaction is short-range, the two-body relative wave-function can deform to avoid the cost of the two particle interactions, a deformation that costs an energy of $E_{\rm F}$. In contrast, the ordering dynamics of the cavity-fermi gas system is not only \textit{captured} by mean-field theory, it renders the (dynamical) idiosyncrasies of mean-field theory \textit{physically manifest}: The fermions each individually react to the \textit{collectively} determined mean-field that is generated by cavity photon mediated all-to-all interactions. Experiments to verify whether effects beyond our present theory set a limit on the rate are feasible and should allow us to firmly elucidate qualitative differences between non-equilibrium dynamics in quantum many-body systems with short and long-range interaction. 

\textit{\textbf{Acknowledgements} ---}We acknowledge stimulating interactions with H. Wu and A. Gomez-Salvador. 
The ETH group acknowledges funding from the Swiss National Science Foundation project 200021\_212899, the Swiss State Secretariat for Education, Research and Innovation (contract number UeM019-1), and NCCR SPIN. E.D. acknowledges funding from ARO grant number W911NF-21-1-0184.
S.C. is grateful for support from the NSF under Grant No. DGE-1845298 \& for the hospitality of the Institute of Theoretical Physics at ETH, Zürich.  J.P.B. and T.Z. acknowledge funding from the Swiss State Secretariat for Education, Research and Innovation (Grants No. MB22.00063 and UeM019-5.1). T.G. and C.-M.H.~acknowledge support by the Swiss National Science Foundation under Division II grants 200020-188687 and 200020-219400, and would like to thank the Institut Henri Poincaré (UAR 839 CNRS-Sorbonne Université) and the LabEx CARMIN (ANR-10-LABX-59-01) for their support. C.-M.H.~acknowledges support in part by grant NSF PHY-1748958 to the Kavli Institute for Theoretical Physics (KITP).S.B.J. acknowledges support from the Deutsche Forschungsgemeinschaft (DFG): Projects A4 and A5 in SFB/Transregio 185: “OSCAR”.


%%%%%%%%%%%%%%%%%%%%%%%%%%%%%%%%%%%%%%%%%%%%%%%%%%%
\onecolumngrid
\appendix
\newpage
\begin{center}
	\textbf{\Large Supplementary Materials}
\end{center}
\normalsize

\setcounter{equation}{0}
\setcounter{figure}{0}
\setcounter{table}{0}
\makeatletter
\setlength\tabcolsep{10pt}
\setcounter{secnumdepth}{2}

%%%%%%%%%%%%%%%%%%%%%%%%%%%%%%%%%%%%%%%%%%%%%%%%%

\section{Self-consistency equation for free fermions}\label{app_free_fermions}

We derive the self-consistency equation (\ref{eq:rate_equation}) for free fermions coupled to a cavity mode with the Hamiltonian (\ref{eq:hamiltonian}) where $\hat{H}_F=\sum_{p\sigma}\epsilon_p \hat{c}^\dag_{p,\sigma}\hat{c}_{p,\sigma}$ is the free fermionic Hamiltonian. We will do so by linearizing the Heisenberg equations of motion $\frac{\mathrm{d} \hat{O}}{\mathrm{d}t} = i \comm{\hat{H}}{\hat{O}}$ for the operator vector $\hat{O}=(\hat{x},\, \hat{p},\, \hat{\rho}_{Q},\, \hat{\rho}_{-Q})^T$. The analysis can easily be extended to the case with multiple ordering wave vectors $Q$ by adding more rows in $\hat{O}$. We adopt the standard notation 
\begin{equation}
    \hat{\rho}_{p,q,\sigma} = \hat{c}^\dag_{p+q,\sigma} \hat{c}_{p,\sigma} \quad \hat{\rho}_{q,\sigma} = \frac{1}{\sqrt{N}} \sum_p \hat{\rho}_{p,q,\sigma} \quad \hat{\rho}_{p,q} = \sum\limits_{\sigma}\hat{\rho}_{p,q,\sigma}.
\end{equation}

For clarity, we stress the difference between operators $\hat{\rho}$ and expectation values $\rho$ in this appendix. By applying the identity $\comm{\hat{c}^\dag_{p1} \hat{c}_{p2}}{\hat{c}^\dag_{p3} \hat{c}_{p4}} = \hat{c}^\dag_{p1} \hat{c}_{p4} \delta_{p2,p3} - \hat{c}^\dag_{p3}\hat{c}_{p2} \delta_{p1,p4}$ for general momenta indices $p_i$, we obtain the commutators
\begin{eqnarray}
    \comm{\hat{H}_F}{\hat{\rho}_{p,q}} &=& (\epsilon_{p+q} - \epsilon_p) \hat{\rho}_{p,q}\\
    \comm{\hat{H}^{\rm int}}{\hat{\rho}_{p,q}} &=& \Lambda \hat{x} \left( \hat{\rho}_{p,q+Q} - \sum_{\sigma}\hat{c}^\dag_{p+q,\sigma}\hat{c}_{p-Q,\sigma}  + \hat{\rho}_{p,q-Q} - \sum_{\sigma}\hat{c}^\dag_{p+q,\sigma} \hat{c}_{p+Q,\sigma}\right)\label{comm_rhopq1}
\end{eqnarray}

Since we are linearizing about the ground state in which the fermions are in a translationally invariant fermi sea and the cavity photon is in a vacuum state, the post-quench expectation value of $\hat{O}$ is small for short times. We should take expectation values in the pre-quench state so that all terms are of the same order. Thereafter (\ref{comm_rhopq1}) reads
\begin{equation}\label{comm_rhopq2}
    \comm{\hat{H}^{\rm int}}{\hat{\rho}_{p,q}}_{\textit{\textrm{lin}}} = (\delta_{q, Q}+\delta_{q, -Q}) \Lambda \hat{x} \sum_{\sigma}\left(n_{p,\sigma} -n_{p+q,\sigma}\right)
\end{equation}
where $n_{p,\sigma}$ is the number of fermions with momentum $p$ and spin $\sigma$ in the fermi sea. For the Bosonic operators, we have
\begin{equation}
    \comm{\hat{H}}{\hat{x}} = -i\hat{p}\Delta_c \quad \comm{\hat{H}}{\hat{p}} = i \big(\Delta_c \hat{x} + \sqrt{N}\Lambda\sum_\sigma (\hat{\rho}_{Q,\sigma} + \hat{\rho}_{Q,\sigma})\big).
    \label{comm_photons}
\end{equation}

For eigenmodes of the system, all the operator dynamics will evolve with the same frequency $\omega$ that is real for oscillatory modes and complex for exponentially growing or decaying modes. In Fourier space, the equation of motion for $\hat{\rho}_{p,q}$ becomes
\begin{equation}
    \omega \hat{\rho}_{p,q} = (\epsilon_{p+q} - \epsilon_p) \hat{\rho}_{p,q} + (\delta_{q, Q}+\delta_{q, -Q}) \Lambda \hat{x} \sum_{\sigma}\left(n_{p,\sigma} -n_{p+q,\sigma}\right)
\end{equation}
which we can sum over $p$ to obtain
\begin{equation}
    \hat{\rho}_{\pm Q} = 2\sqrt{N}\Lambda \hat{x}\frac{1}{2N}\sum_{p,\sigma}\frac{n_{p\pm Q,\sigma}-n_{p,\sigma}}{\epsilon_{p\pm Q} - \epsilon_p - \omega} = -2\sqrt{N}\Lambda\hat{x}\chi(\pm Q,\omega)
\end{equation}
where we recognize the Lindhard response function at momentum $\pm Q$ and frequency $\omega$. The factor of 2 in the denominator compensates for the sum over spin which is usually not in the definition of the Lindhard function. We have thus derived the proportionality between $\hat{x}$ and $\hat{\rho}_Q$ that was sketched in the main text. For all other momenta $q\neq\pm Q$, the equations of motion of $\hat{\rho}_q$ are decoupled from the dynamics of $\hat{\rho}_{\pm Q}$. We see that in Fourier space, the equations of motion of $\hat{O}$ are closed while in the time domain, they were not. The final set of equations
\begin{eqnarray}
    0 &=& i\omega\hat{x} - \Delta_C \hat{p}\\
    0 &=& \Delta_C\hat{x} + i\omega\hat{p} +\sqrt{N}\Lambda\sum_Q(\hat{\rho}_{Q}+\hat{\rho}_{-Q})\\
    0 &=& 4\sqrt{N}\Lambda\sum_Q\chi(Q,\omega)\hat{x} + \sum_Q(\hat{\rho}_{Q}+\hat{\rho}_{-Q})
\end{eqnarray}
can be shown to have non-trivial solutions when the condition
\begin{equation}\label{app1_det}
    \omega^2 = \Delta_C^2 - 4\Lambda^2 N\Delta_C\sum_{Q}\chi(Q,\omega)
\end{equation}
holds. We recognize this as the self-consistency equation (\ref{eq:rate_equation}). We emphasize the fermionic susceptibility in the self-consistency equation depends only on the pre-quench fermionic state.

%%%%%%%%%%%%%%%%%%%%%%%%%%%%%%%%%%%%%%%%%%%%%%%%%%%%
\section{Response functions}\label{app_response_functions}
The purpose of this appendix is to derive from (\ref{eq:rate_equation}) the more sophisticated (\ref{eq:rate}) and point to features of density-density response functions useful to our calculations. Since we want to compute instability modes with imaginary frequencies but the response functions found in literature are usually the retarded response functions defined only for real $\omega$, we use the Lehman representation of $\chi$ \cite{dupuis2023field}. By rescaling $z=\frac{\omega}{\Delta_C}=z'+iz''$, the self-consistency equation (\ref{eq:rate_equation}) becomes
\begin{equation}\label{eq:app_modi}
    z^2 = 1 - \frac{4\Lambda^2 N}{\Delta_c} \int_{-\infty}^\infty \frac{\mathrm{d}\omega}{\pi} \frac{\chi''(\omega)}{\omega - \Delta_c z},
\end{equation}
with  $\chi''(\omega) = \Im(\chi^R(Q,\omega))$ where $\chi^R(Q,\omega)$ is the standard causal response function of fermions at wave vector $Q$ and frequency $\omega$. In the Fourier convention that we are using, $z''<0$ corresponds to exponential growth. The next step is to separate equation (\ref{eq:app_modi}) into real and imaginary parts
\begin{equation}
\begin{split}
z'^2 - z''^2 &= 1 - \frac{4\Lambda^2 N}{\Delta_c} \int\frac{\mathrm{d}\omega}{\pi} \frac{\chi''(\omega
)}{(\omega - \Delta_c z')^2 + \Delta_c^2  z''^2} (\omega - \Delta_c  z'),\\
2 z' z'' &= -\frac{4\Lambda^2 N}{\Delta_c} \int\frac{\mathrm{d}\omega}{\pi} \frac{\chi''(\omega)}{(\omega - \Delta_c  z')^2 + \Delta_c^2  z''^2} \Delta_c  z''.
\label{eqn:self-consistency}
\end{split}
\end{equation}

We recall that $\chi''(\omega)$ is an odd function. Therefore equations (\ref{eqn:self-consistency}) admit imaginary solutions $z'=0,\,z''\neq0$ that satisfy
\begin{equation}\label{eq:app_imag_self1}
1 + z''^2 = \frac{4 \Lambda^2 N}{\Delta_c} \int_{-\infty}^\infty \frac{\mathrm{d} \omega}{\pi}\frac{\omega \Im(\chi^R(Q,\omega))}{\omega^2 + \Delta_c^2 z''^2}.
\end{equation}
which is precisely (\ref{eq:rate}) upon substituting $z''=\Gamma$. It can similarly be shown there exist purely real solutions which lie in the vicinity of the photon detuning. Since we are looking for the strongest instability of the system, one may worry that a complex solution with a larger imaginary part also exists. We checked the system of equations (\ref{eqn:self-consistency}) numerically and found no such solutions for our parameter regimes. We always found 4 eigenmodes; 2 real ones and two that are imaginary above the superradiant threshold and real below.

%%%%%%%%%%%%%%%%%%%%%%%%%%%%%%%%%%%%%%%%%%%%%%%%%%%%
\section{Adding dissipation}\label{app_dissipation} 
Dissipation effects are both an inevitable and interesting effect in cavity research. In this appendix, we show that our analysis works well even with the addition of dissipation. We include  dissipation in the way known from input-output theory \cite{inputoutput}. The modified self-consistency equation is
\begin{equation}
    (\omega-i\kappa)^2 = \Delta_C^2 - 4\Delta_C N\Lambda^2\sum_Q\chi(Q,\omega)
\end{equation}
where $\kappa$ is the photon leakage rate. Following the same arguments as for the closed system, there exist purely imaginary solutions to the self-consistency equation. (\ref{eq:app_imag_self1}) is modified to
\begin{equation}\label{eq:app_imag_self2}
(z''-\frac{\kappa}{\Delta_C})^2 + 1 = \frac{4 \Lambda^2 N}{\Delta_c} \int_{-\infty}^\infty \frac{\mathrm{d} \omega}{\pi}\frac{\omega \Im(\chi^R(Q,\omega))}{\omega^2 + \Delta_c^2 z''^2}
\end{equation}
which has a superradiant threshold ${\Lambda}_C^2(\kappa) = \Lambda_C^2(\kappa=0)\left(1+(\frac{\kappa}{\Delta_C})^2 \right)$. Below the threshold, all solutions have a decaying component. Notice that the fermionic response function is the same as in the closed system as we only included photon loss.

%%%%%%%%%%%%%%%%%%%%%%%%%%%%%%%%%%%%%%%%%%%%%%%%%%%%
\section{Adding interactions}
While the derivation in appendix \ref{app_free_fermions} holds only for free fermions, the final self-consistency equation is more general. The analysis in \ref{app_response_functions} and the corresponding eigenmode equation \eqref{eq:rate_equation} also hold for any system coupled to cavity photons, regardless of the microscopic details, if we simply generalize the density-density response function to correspond to the specific system of interest. To prove this,  we map the derivation onto a more general problem of computing the density response function and take advantage of the results already existing in literature.  
We will rely on two key facts we observe in the free fermion equations. First, considering equation \eqref{comm_photons}, we notice that to close the system of equations \eqref{comm_rhopq1} and \eqref{comm_photons}, we only need to compute the density operator $\hat{\rho}_Q$ up to linear order in $\hat{x}$. Second, when computing equations of motion, after computing the commutators it is common to take expectation values up to linear order (as described in \ref{app_free_fermions}). Therefore, we see that from perspective of fermionic operators, there is no difference between taking expectation value of $\hat{x}$ before, or after computing the commutators. This is key, as we see that then the original interaction term can be re-written as
\begin{equation}
    \hat{H}^{int} = \Lambda \sqrt{N}(\expval{\hat{x}} \hat{\rho}_Q + \hat{x} \expval{\hat{\rho}_Q}).
\end{equation}

Looking at the part acting on fermions, we can recognize the canonical linear response-type density driving $\phi_Q \hat{\rho}_Q$, where the usual field $\phi_Q$ has been replaced by $\Lambda \sqrt{N} \expval{\hat{x}}$. Therefore, following the usual response theory \cite{dupuis2023field}
\begin{equation}
    \expval{\hat{\rho}_Q(Q,\omega)} = -\chi(Q,\omega) \Lambda \sqrt{N} \expval{\hat{x}(\omega)}
\end{equation}

with the equality holding for every $Q$ component individually, up to linear order. Therefore, equations of motion hold regardless of the microscopic details, we should just use the appropriate response function. \\

\subsection{RPA analysis}

In this work we consider a system of fermions having a contact interaction $U$, such that the fermionic Hamiltonian can be written
\begin{equation}
    \hat{H}_F = \sum_k \xi_k \hat{c}^\dag_k \hat{c}_k + U \sum_{k,k',q,\sigma,\sigma'} \hat{c}^\dag_{k+q, \sigma} \hat{c}^\dag_{k'-q,\sigma'} \hat{c}_{k',\sigma'} \hat{c}_{k,\sigma}
\end{equation}

where $\xi_k = k^2/2m - \mu$. We model the ground state of the system using the usual BCS theory and focus on computing the response of the system within the RPA approximation, which is known to qualitatively capture particle hole continuum as well as the sound mode \cite{Response-sound}. Before we proceed, we note that $U$ is the interaction which should be adjusted, along with the high momentum cut-off $k_M$, to give the correct scattering length

\begin{equation}
    \frac{m}{4\pi a_s} = \frac{1}{U} + \frac{1}{V} \sum_k^{k_M} \frac{m}{k^2}.
    \label{eqn:scattering_length}
\end{equation}

To obtain the physical quantities, we consider the simultaneous limit of $U \rightarrow 0, k_M \rightarrow \infty$  with $a_s$ fixed. Within the BCS theory, the gap equation becomes
\begin{equation}
    -\frac{1}{U} = \frac{1}{V} \sum_k^{k_M} \frac{1}{E_k}
\end{equation}

where $E_k = \sqrt{\xi_k^2 + |\Delta|^2}$ and $\Delta$ is the gap of the system. The gap equation can be regularised by subtracting the scattering length equation \eqref{eqn:scattering_length} and by letting the cut-off $k_M \rightarrow \infty$.
Now that we have discussed the ground state properties and regularisation, let us turn to computing the response of the system. Following previous work, within the RPA approximation, the response can be written as \cite{RPA,RPA-FT,RPA-weak}

\begin{equation}
    \chi^{RPA} = \frac{\chi^0}{1 - U \chi^0 (\sigma_x \oplus \sigma_x)}
\end{equation}

where  $\chi_0$ is a $4 \times 4$ response matrix acting on $(\hat{\rho}_{Q,\uparrow},\, \hat{\rho}_{Q,\downarrow}, \,\hat{\eta}_Q, \,\hat{\eta}_Q^\dag)^T$, where $\hat{\eta}^\dag_q = \sum_p \hat{c}^\dag_{p+q\uparrow}\hat{c}^\dag_{-p}$. We assume that the above expression is properly regularized and the limits $U\rightarrow 0, k_c \rightarrow \infty $ are taken appropriately.  
Individually, each $\chi^0$ can therefore be derived by computing the appropriate equations of motion using the BCS Hamiltonian and by expressing the ground state occupations.  We can show that for zero temperature and translationally invariant ground state, the entries in the $\chi^0$ matrix can be related $\chi^0_{11} = \chi^0_{22}$, $\chi^0_{12} = \chi^{0}_{21} = - \chi^0_{33} = - \chi^{0}_{44}$, $\chi^0_{31} = \chi^0_{32} = \chi^0_{14} = \chi^0_{24} $, $\chi^0_{41} = \chi^0_{42} = \chi^0_{13} = \chi^0_{23}$. All the elements are convergent as the cut-off $k_c \rightarrow \infty$ except $\chi^0_{43}$ and $\chi^0_{34}$. Therefore we define the regularised counterpart of the integrals $\Tilde{\chi}_{43} = \chi^0_{43} - 1/U$ and $\Tilde{\chi}^0_{34} = \chi^0_{34} -1/U$.
In limit $U \rightarrow 0 $, density response function reduces to

\begin{equation}
    \chi^{RPA} = 2(\chi^0_{11} + \chi^0_{12}) - 4 \frac{(\chi^0_{31})^2 \Tilde{\chi}_{43} + (\chi^0_{41})^2 \Tilde{\chi}_34 + 2\chi^0_{12}\chi^0_{31}\chi^0_{41}}{\Tilde{\chi}_{34}\Tilde{\chi}_{43} - (\chi^0_{12})^2} 
\end{equation}

where the individual coefficients are given \cite{RPA}

\begin{eqnarray}
    &\chi^0_{11} = \frac{1}{4} \sum_k 1 - \frac{\xi_k \xi_{k+q}}{E_k E_{k+q}} F_1(k,q) \\
    &\chi^0_{12} = \frac{1}{4} \sum_k \frac{|\Delta|^2}{E_k E_{k+q}} F_1(k,q) \\
    &\chi^0_{31} = \frac{1}{8} \sum_k \frac{\Delta}{E_k E_{k+q}} \left( (\xi_k + \xi_{k+q})F_1(k,q) + (E_k + E_{k+1}) F_2(k,q)\right)\\
    &\chi^0_{41} = \frac{1}{8}  \sum_k \frac{\Delta}{E_k E_{k+q}} \left( (\xi_k + \xi_{k+q})F_1(k,q) - (E_k + E_{k+1}) F_2(k,q)\right)\\
    &\chi^0_{43} = \frac{1}{4} \sum_k \left( \left( 1 + \frac{\xi_k \xi_{k+q}}{E_k E_{k+1}}\right) F_1(k,q) - \left( \frac{\xi_k}{E_k} + \frac{\xi_{k+q}}{E_{k+q}} \right) F_2(k,q) \right) \\
    &\chi^0_{34} = \frac{1}{4} \sum_k \left( \left( 1 + \frac{\xi_k \xi_{k+q}}{E_k E_{k+1}}\right) F_1(k,q) + \left( \frac{\xi_k}{E_k} - \frac{\xi_{k+q}}{E_{k+q}} \right) F_2(k,q) \right)\\
\end{eqnarray}

where $F_{1/2} = \frac{1}{i\omega_n - (E_{k} + E_{k+q})} \mp  \frac{1}{i\omega_n + (E_{k} + E_{k+q})}$ are the usual pole functions, which need to be analytically continued $i\omega_n \rightarrow \omega + i\eta$ to obtain the retarded response function at frequency $\omega$.

\section{Comparing with experimental data}
\label{sec:Exp-comp}

Parameters used to compare to lossy free fermions are in Fig \ref{fig:Fig3}: fermion number per spin
$N = 10^5$, trap frequencies $(\omega_x,\omega_y,\omega_z)/2\pi = (290,380,85) \text{Hz}$, effective pump-cavity detuning $\Delta_c/2\pi = 720 \text{kHz}$, decay rate $\kappa/2\pi = 677 \text{kHz}$, ordering wavevector $|\textbf{k}_+| = |\textbf{k}_-|= 2.03 k_{\rm F}$ and we can compute the Fermi energy using $E_{\rm F} = (\omega_x \omega_y \omega_z)^{1/3} (6 N)^{1/3}$.

%%%%%%%%%%%%%%%%%%%%%%%%%%%%%%%%%%%%%%%%%%%%%%%%%%%%
\section{Computing the pre-factor for free fermions}
\label{sec:FF-pre-factor}
The goal of this appendix is to comprehensively derive the photon and density wave order parameter pre-factors for the unstable mode of the free fermion gas coupled to a cavity. In the next appendix, we will show how to generalize the procedure to interacting fermions. We will neglect dissipation. The procedure could be used to compute many types of instability pre-factors and is thus worth to write down in detail. We will again start from the linearized equations of motion 
\begin{equation}\label{app_linear}
    \frac{\mathrm{d}}{\mathrm{d}t}\hat{O}(t) = M\hat{O}(t).
\end{equation}

However, $\hat{O} = (\hat{x},\,\hat{p},\,\hat{\rho}_{p,Q},\,\hat{\rho}_{p,-Q})^T$ is now a vector of length $2+2Nn_Q$ where $N$ is the number of atoms and $n_Q$ is the number of ordering wave vectors. Like $\chi$, the matrix $M$ depends only on the initial state of the system. We compute its entries from equations of motion in appendix \ref{app_free_fermions}. We see that to get a closed system of linear equations in the time domain, we had to extend the operator $\hat{O}$. However, we did not need to include operators $\hat{\rho}_{p,q}$ for $q\neq\pm Q$.
We take the Laplace transform of (\ref{app_linear}) to obtain
\begin{equation}
    \hat{\Tilde{O}}(s) = \frac{1}{s-M} \hat{O}(t=0)
\end{equation}
where the Laplace transformation is done w.r.t. the complex variable $s$ and we denote the transformed vector by a tilde. Now that we have placed all the $s$ dependence on the RHS, we can invert the transformation by a Bromwich integral
\begin{equation}
\hat{O}(t) = \frac{1}{2\pi i } \lim_{T \rightarrow \infty} \int_{\gamma - iT}^{\gamma + iT}\mathrm{d}s  e^{st} \frac{1}{s-M} \hat{O}(0)
\end{equation}
where $\gamma\in\mathbb{R}$ is chosen so that of all the poles of $1/(s-M)$ lie on the left side of the vertical integration line. Since $e^{st}$ is an exponentially decaying function on the left complex half-plane, we can close the integration contour with a semicircle at infinity and apply Residue Theorem to compute the integral. Up to here, our equation was exact, and in principle we should include all the eigenmodes that will show up as poles of $\frac{1}{s-M}$, as we will see further in the derivation. Since we want to compute the pre-factor of the dominant exponentially growing mode, we consider only the pole at $s=\Gamma$
\begin{equation}
    \hat{O}(t) = e^{\Gamma t} \lim_{s \rightarrow \Gamma}\left(\frac{s - \Gamma}{s - M}\right) \hat{O}(0).
\end{equation}

A comment on why we chose to use Laplace instead of Fourier transforms is in place. If we had done a Fourier transform, we could not have closed the contour around the relevant pole when going back to the time domain as the function $e^{i\omega t}$ diverges in that half-plane. Furthermore, a Fourier transform of an exponentially diverging function is not well defined while its Laplace transform is well-behaved. These details did not ail the computation of the self-consistency equation because we did not need the eigenvectors but only the eigenvalues of the exponentially growing modes. We will see the Laplace transform yields the same self-consistency equation. Thus, it is a natural means to analyzing an unstable mode.

We now go through the mathematical details of computing the residue of
\begin{equation}
        s - M = \begin{pmatrix}
            s                  & -\Delta_C & 0        & 0 & \cdots\\
            \Delta_C                 & s   & \Lambda & \Lambda & \cdots \\
            -i\Lambda (n_p - n_{p+Q})  & 0         & s + i(\epsilon_p - \epsilon_{p+Q})       & 0       & \cdots \\
            \vdots & 0 & 0 & \ddots \\ 
        \end{pmatrix}
\end{equation}
in which the momenta $p$ run over a fictional discretized box grid at half filling so the number of momenta is equal to the number of atoms. Each ordering wave vector contributes $2N$ rows and columns to the matrix, $N$ for $Q$ and $N$ for $-Q$. For clarity, we will work with only one $Q$ until the end result and then comment on how to include multiple wave vectors. We can separate the matrix into two parts
\begin{equation}
    s-M = \begin{pmatrix}
        1 & 0& 0& \cdots\\ 
        0 & 1 & 0 & \cdots \\
        0 & 0 & s + i(\epsilon_p - \epsilon_{p+Q}) & \cdots\\
        0 & 0 & 0 & \ddots
    \end{pmatrix} \begin{pmatrix}
            s                  & -\Delta_C & 0        & 0        & \cdots\\
            \Delta_C           & s         & \Lambda  &  \Lambda & \cdots \\
            \frac{-i\Lambda (n_p - n_{p+Q})}{s + i(\epsilon_p - \epsilon_{p+Q}) }  & 0         & 1      & 0       & \cdots \\
            \vdots & 0 & 0 & \ddots \\ 
    \end{pmatrix}.
\end{equation}

 We label the first matrix $A$. We wil perform a unitary change of basis on the second matrix with 
\begin{equation}
    U = \begin{pmatrix}
        1_2 & 0 & 0\\
        0 & U_N & 0\\
        0 & 0 & U_N
    \end{pmatrix}
\end{equation}
where the $N\times N$ matrix $U_N$ has elements $(U_N)_{nm} = \frac{1}{\sqrt{N}} e^{\frac{2\pi i}{N} n m}$ with $0\leq n,m<N$. After the unitary transform
\begin{equation}
    s - M = A U^\dag \begin{pmatrix}
            s                  & -\Delta_C & 0        & 0 &  0 &\cdots\\
            \Delta_C                 & s   & \sqrt{N}\Lambda & 0 &  0 &\cdots \\
            2\sqrt{N}\Lambda\chi(Q,-is)  & 0         & 1      & 0    & 0  & \cdots \\
            2\sqrt{N}\Lambda\chi^{(1)}(Q,-is) & 0 & 0 & 1 & \cdots\\
            \vdots & 0 & 0 & 0 & \ddots \\ 
    \end{pmatrix} U 
\end{equation}
we get a matrix that has a filled first column, a non-zero $3\times 3$ first block, a diagonal of ones, and zeroes everywhere else. This structure will be essential for the simplifications we will make next. $\chi=\chi^{(0)}$ is the usual density-density response function and
\begin{equation}
    \chi^{(m)}(Q,\omega) = \frac{1}{2N}\sum_{p,\sigma}\frac{n_{p,\sigma} - n_{p+Q,\sigma}}{\epsilon_{p+Q} - \epsilon_p - \omega}e^{i2\pi \frac{pm}{N}}
\end{equation}
are response functions for drives that have asymmetrical sums of $\rho_{pq}$. We will see they will not enter the final equations. Since the response function $\chi$ is invariant under (complex) time reversal, we can write $\chi(Q, is)$ instead of $\chi(Q,-is)$. We can shuffle the matrix so that 
\begin{equation}\label{app2_Mbar}
    s - M = A U^\dag \begin{pmatrix}
            s                  & -\Delta_C & 0        & 0 &  0 &\cdots\\
            \Delta_C                 & s   & \sqrt{N}\Lambda & \sqrt{N}\Lambda &  0 &\cdots \\
            2\sqrt{N}\Lambda\chi(Q,is)  & 0         & 1      & 0    & 0  & \cdots \\
            2\sqrt{N}\Lambda\chi(-Q,is) & 0 & 0 & 1 & \cdots\\
            \vdots & 0 & 0 & 0 & \ddots \\ 
    \end{pmatrix} U
\end{equation}
but we have to be consistent in shuffling terms in $\hat{O}$ as well. So far, we have shown
\begin{equation}
    s - M = A(s) U^\dag \bar{M}(s) U
\end{equation}
where $\bar{M}(s)$ is the matrix in (\ref{app2_Mbar}). It's easy to see

\begin{equation}
    \frac{1}{s-M} = U^\dag \bar{M}(s)^{-1} U A^{-1}(s).
\end{equation}

We are interested in the residue of this matrix at the pole $s=\Gamma$ corresponding to the exponentially growing mode. If the pole is to be simple, it can only come from one source and this is the matrix $\bar{M}$. More specifically, the pole comes from the determinant $\det(\bar{M})$. Therefore, the time dependence of $\hat{O}$ is
\begin{eqnarray}
    \hat{O}(t) = e^{\Gamma t} \textit{Res} \left( \frac{1}{\det(\bar{M}(s))} \right)_{s = \Gamma}  U^\dag \big(\det(\bar{M}(\Gamma))\bar{M}^{-1}(\Gamma) \big) U A^{-1}(\Gamma) \hat{O}(t = 0)
\end{eqnarray}
where, given the structure, $\det(\bar{M}(\Gamma))\bar{M}^{-1}(\Gamma)$ is a well defined object at $\Gamma$. In fact, it can be computed analytically 
\begin{equation}\label{app_Mbar_inv}
    \det(\bar{M}(\Gamma))\bar{M}^{-1}(\Gamma) = \begin{pmatrix}
        \Gamma & \Delta_c & -\sqrt{N} \Delta_c \Lambda & -\sqrt{N} \Delta_c \Lambda & 0 &\cdots \\
        \frac{\Gamma^2}{\Delta_c} & \Gamma & -\Lambda\sqrt{N} \Gamma & -\Lambda\sqrt{N} \Gamma & 0 & \cdots \\
        -2\Lambda\sqrt{N} \chi\Gamma &  -2\sqrt{N} \Delta_c \Lambda \chi & 2N \Delta_c \Lambda^2 \chi & 2N \Delta_c \Lambda^2 \chi & 0 & \cdots \\
        -2\Lambda\sqrt{N} \chi\Gamma &  -2\sqrt{N} \Delta_c \Lambda \chi & 2N \Delta_c \Lambda^2 \chi & 2N \Delta_c \Lambda^2 \chi & 0 & \cdots \\
        \vdots & \vdots & \vdots & 0 & 0 & \cdots \\
    \end{pmatrix}
\end{equation}
with shorthand notation $\chi\equiv\chi(Q,i\Gamma)$. The block structure of $\det(\bar{M}(\Gamma))\bar{M}^{-1}(\Gamma)$ with zeroes everywhere from the fifth column onwards will greatly simplify the analysis.

We now have to work through two tasks: first, to compute the residue and show the pole of the matrix corresponds to the unstable mode, and then to compute the fluctuations in the ground state that seed the exponential growth.

Due to the block structure of $\bar{M}(s)$, the determinant is easy to compute:
\begin{equation}
    \mathrm{det}\left(\bar{M}(s)\right) = \Delta_c^2 - 4N \Lambda^2 \Delta_c \chi(Q,is)  + s^2
\end{equation}

We can see that for $s=i\omega$, we get the same self-consistency equation as before. At $s=\Gamma$, the residue can be computed by L'Hospital's rule
\begin{equation}\label{eq:app_residue}
\textit{Res} \left( \frac{1}{\det(\bar{M}(s))} \right)_{s = \Gamma} = \frac{1}{2\Gamma - 4 N\Lambda^2 \Delta_c \left. \frac{\mathrm{d}\chi(Q,is)}{\mathrm{d}s}\right|_\Gamma} = \frac{1}{2\Gamma} \cdot \frac{1}{1 + 4N\Lambda^2 \Delta_c \int_{-\infty}^\infty \frac{\mathrm{d}\omega}{\pi} \frac{\Im \chi(Q, \omega) \omega}{(\omega^2 + \Gamma^2)^2})}.
\end{equation}

We used the spectral representation of $\chi$ to get the second equality. Again, (\ref{eq:app_residue}) is a good result in that we only need the knowledge of the ground state properties and the instability rate to compute the residue.

The expectation value of $\hat{O}(t)$ will be zero at all times as $\hat{O}(t=0)=0$ in the ground state. What we are really interested in are expectation values
\begin{equation}
    \expval{\hat{O}_i \hat{O}_j}(t) = e^{2\Gamma t} \left(\textit{Res} \left( \frac{1}{\det(\bar{M}(s))} \right)_{s = \Gamma} \right)^2 \sum_j(\det(\bar{M}(\Gamma))\bar{M}^{-1}(\Gamma) )_{ik} (\det(\bar{M}(\Gamma))\bar{M}^{-1}(\Gamma) )_{jl} \expval{{\hat{O}'}_k \hat{O}'_l}_0
\end{equation}
with the transformed and reshuffled vector
\begin{equation}
    \hat{O}'= U A^{-1} \hat{O} = U\begin{pmatrix}
        \hat{x} \\ \hat{p} \\ \frac{\hat{\rho}_{pQ}}{\Gamma + i(\epsilon_p - \epsilon_{p+Q})} \\ \frac{\hat{\rho}_{p-Q}}{\Gamma + i(\epsilon_p - \epsilon_{p-Q})} \\ \vdots
    \end{pmatrix} = \begin{pmatrix}
        \hat{x} \\ \hat{p} \\ \frac{1}{\sqrt{N}}\sum_p \frac{\hat{\rho}_{pQ}}{\Gamma + i(\epsilon_p - \epsilon_{p+Q})} \\ \frac{1}{\sqrt{N}}\sum_p \frac{\hat{\rho}_{p,-Q}}{\Gamma + i(\epsilon_p - \epsilon_{p,-Q})}\\ \vdots
    \end{pmatrix}.
\end{equation}

We see we need expectation values of quadratures in the ground state, denoted by the subscript $0$. For the photon, it's $\expval{\hat{x}^2}_0=\frac{1}{2}$ and $\expval{\hat{p}^2}_0=\frac{1}{2}$ while all cross terms and photon-fermion expectation values are 0. However, they are  dominated by the fermionic density fluctuations
\begin{equation}
    \left\langle \frac{1}{N}\sum_{k,p}\frac{ \hat{\rho}_{pQ}\hat{\rho}_{k,-Q}}{(\Gamma + i(\epsilon_p-\epsilon_{p+Q}))(\Gamma + i(\epsilon_k - \epsilon_{k-Q}))}\right\rangle_0 = \frac{2}{N}\sum_p \frac{(1-n_{p+Q,\sigma})n_{p,\sigma}}{\Gamma^2 + (\epsilon_{p+Q} - \epsilon_{p})^2} = 2\int_0^\infty \frac{\mathrm{d}\omega}{\pi}\frac{\chi''(\omega)}{\omega^2 + \Gamma^2}
\end{equation}
and we will ignore them from now on. We obtained the last equality by inserting $\chi''(\omega,Q)= \frac{\pi}{N}\sum_{p}n_{p,\sigma}(1-n_{p+Q,\sigma})\left(\delta(\epsilon_{p+Q} - \epsilon_p - \omega) - \delta(\epsilon_{p+Q} - \epsilon_p + \omega)\right)$. Somewhat surprisingly, the fermionic noise is not only the static structure factor at $Q$. We see that even though the drive is a symmetric sum of $\hat{\rho}_{pq}$, the fluctuations seeding the instability are more complex. 

Since $\det(\bar{M}(\Gamma))\bar{M}^{-1}(\Gamma)$ has only zeroes after the fourth column, the quadratures shown above alone determine the pre-factor. We are ready to compute the photon number pre-factor. Reading off the relevant matrix elements from (\ref{app_Mbar_inv}), we arrive at
\begin{equation}\label{app2_seedn1}
    \expval{\hat{n}}(t)\approx \frac{1}{2}\expval{\hat{x}^2+\hat{p}^2}(t) \approx e^{2\Gamma t} \frac{\Gamma^2 + \Delta_C^2}{8\Gamma^2} \cdot \left(\frac{1}{1 + 4N\Lambda^2 \Delta_c \int_{-\infty}^\infty \frac{\mathrm{d}\omega}{\pi}\frac{\chi''(\omega) \omega}{(\omega^2 + \Gamma^2)^2}}\right)^2 N \Lambda^2  4\int_0^\infty\frac{\mathrm{d}\omega}{\pi} \frac{\chi''(\omega)}{\omega^2 + \Gamma^2}
\end{equation}
which becomes (\ref{eq:seed}) upon substituting the expression for critical light-matter coupling strength. The ratio of the $p$ and $x$ contributions is $\frac{\Gamma}{\Delta_C}$ which is negligible near the phase transition but becomes significant at extremely deep quenches. 

%Now is a time to remind that while the seed was defined as the prefactor of the exponentially growing term, the total time dependency of the photon number is exponential only at intermediate times when the other modes are relatively weak but the system still behaves linearly. At time $t=0$ expression (\ref{app2_seedn1}) clearly does not yield $\expval{n}=0$ as it should for a quench from a photon vacuum state. This is not due to the fact we neglected contributions from photon quadratures and did no subtract $\frac{1}{2}$ but rather because we ignored contributions from all the other poles in the contour integral. To get to \eqref{eq:seed}, we insert definitions of the critical coupling strength. The procedure to obtain the seed for density fluctuations would have been exactly the same except for the matrix elements.

\section{Computing the pre-factor with interactions}

While the derivation in appendix \ref{sec:FF-pre-factor} for the prefactor to the exponential growth holds only in the free fermion case, the final relation is more general. Using linear response theory, we derive in this section the general seed relation for an arbitrary system coupled to a cavity mode.
From our previous calculations, we know the equations of motion for photonic fields are, after linearization
\begin{equation}
    \dot{\hat{x}} = -i\hat{p}\Delta_c \quad \dot{\hat{p}} = i \big(\Delta_c \hat{x} + \sqrt{N}\Lambda\sum_\sigma (\hat{\rho}_{Q,\sigma} + \hat{\rho}_{Q,\sigma})\big).
    \label{comm_photons}
\end{equation}

The above equation in Laplace's transformed form becomes
\begin{equation}
    s\hat{x}(s) - \hat{x}(t = 0) = -i \Delta_c \hat{p}(s) \quad s \hat{p}(s) - \hat{p}(t = 0) = i\Delta_c \hat{x}(s) + i\sqrt{N} \Lambda \hat{\rho}_Q(s)
\end{equation}

Equivalently, the density equation becomes (by using the Kubo formula)
\begin{equation}
    \hat{\rho}_Q(s) = \sqrt{N} \Lambda \chi(Q,s)  \hat{x}(s) + \hat{\rho}^0_Q(s) 
\end{equation}

where we introduced $\hat{\rho}^0(t) = e^{i\hat{H}_{\rm F}t} \hat{\rho}(t = 0 ) e^{-i\hat{H}_{\rm F}t }$ and $\hat{\rho}^0(s)$ is the appropriate Laplace transform. Combining all the equations, we can obtain an equation which only depends on $\hat{x}(s)$

\begin{equation}
    \hat{x}(s) = \frac{1}{D(s)} \left( s \hat{x}(t = 0) + \Delta_c \hat{p}(t = 0) + \sqrt{N} \Lambda \Delta_c \hat{\rho}^0_Q(s)\right)
\end{equation}

Doing the inverse Laplace transform and focusing only on the exponentially increasing solution gives

\begin{equation}\label{app:eq:x_t}
    \hat{x}(t) = e^{\Gamma t} Res\left(\frac{1}{D(s)}\right)_{s\rightarrow \Gamma} \left( \Gamma \hat{x}(t = 0) + \Delta_c \hat{p}(t = 0) + \sqrt{N} \Lambda \Delta_c \hat{\rho}^0_Q(\Gamma)\right)
\end{equation}

We can now turn to compute $\expval{\hat{\rho}^0_Q \hat{\rho}^0_{-Q}}$. To keep the derivation general, we use Lehmann representation, where we assume system's state is given by the density matrix $e^{-\beta \hat{H}_{\rm F}}$ and has eigenstates $\ket{n}$ with energies $E_n$. By using this definition, it follows that in the eigenbasis

\begin{equation}\label{app:eq:rho_basis}
    \hat{\rho}^0_Q(s) = \sum_{nm} \frac{\rho_{nm}}{s + i(E_n - E_m)} \ketbra{n}{m}
\end{equation}

Using this result we can now express the initial pre-factor coming from the density fluctuations

\begin{equation}
    \expval{\hat{\rho}^0_Q(s) \hat{\rho}^0_{-Q}(s)} = \Tr{ e^{-\beta \hat{H}_{\rm F}} \hat{\rho}^0_Q(s) \hat{\rho}^0_{-Q}(s)} = \sum_{nm} e^{-\beta E_n} \frac{\rho^\dag_{nm} \rho_{mn}}{s^2 + (E_m - E_n)^2} = \int_0^\infty \frac{d\omega }{\pi} \frac{\chi''(\omega)}{\omega^2 + s^2} 
\end{equation}

where for the first equality we have used (\ref{app:eq:rho_basis}) and allowed for finite temperature. The second equality follows from the usual definition of the imaginary part of the7 response function 

\begin{equation}
\chi''(\omega) = \pi \sum_{nm} \rho_n \left( \rho_{nm} \rho^\dag_{mn} \delta(\omega - (E_m - E_n)) - \rho^\dag_{nm} \rho_{mn} \delta(\omega + (E_m - E_n)) \right) 
\end{equation}

By squaring the matrix and collecting all the terms, we obtain the same expression as in the previous section, only now the response function is describing an arbitrary system coupled to a cavity.

\end{document}